\documentclass[prl,twocolumn,showpacs,amsmath,amssymb,floatfix,altaffillsymbol]{revtex4-1}
\usepackage{graphicx}
\usepackage{ulem}
\usepackage{soul}
\usepackage{dcolumn}
\usepackage{bm}
\usepackage{amsmath}
\usepackage{latexsym}
\usepackage{amsfonts}
\usepackage{amssymb}
\usepackage{verbatim}
\usepackage[rightcaption]{sidecap}
\usepackage{color}\makeatletter

\newcommand{\Rmnum}[1]{\expandafter\@slowromancap\romannumeral  #1@}

\newcommand{\ch}{\hat{c}}
\newcommand{\chd}{\hat{c}^\dagger}

\newcommand{\Hh}{\hat{H}}

\newcommand{\nh}{\hat{n}}

\makeatother
\begin{document}
\title{Merging Features from GreenÕs Functions and Time Dependent Density Functional
Theory: A Route to the Description of Correlated Materials out of Equilibrium?}
\date{\today}
\author{M. Hopjan}
\affiliation{Mathematical Physics Division, Department of Physics, Lund University, 22100  Lund, Sweden; \\and European Theoretical Spectroscopy Facility (ETSF)}
\author{D. Karlsson}
\altaffiliation[Current address: ]{
Department of Physics,
Nanoscience Center P.O.Box 35 FI-40014 University of Jyv\"askyl\"a, Finland.}
\affiliation{Mathematical Physics Division, Department of Physics, Lund University, 22100  Lund, Sweden; \\and European Theoretical Spectroscopy Facility (ETSF)}
\author{S. Ydman}
\affiliation{Mathematical Physics Division, Department of Physics, Lund University, 22100  Lund, Sweden; \\and European Theoretical Spectroscopy Facility (ETSF)}
\author{C. Verdozzi}
\affiliation{Mathematical Physics Division, Department of Physics, Lund University, 22100  Lund, Sweden; \\and European Theoretical Spectroscopy Facility (ETSF)}
\author{C.-O. Almbladh}
\affiliation{Mathematical Physics Division, Department of Physics, Lund University, 22100  Lund, Sweden; \\and European Theoretical Spectroscopy Facility (ETSF)}
\begin{abstract}
We propose a description of nonequilibrium systems via a simple protocol that combines exchange-correlation
potentials from density functional theory with self-energies of many-body perturbation theory.
The approach, aimed to avoid double counting of interactions, is tested against exact results in Hubbard-type
systems, with respect to interaction strength, perturbation speed and inhomogeneity, and system
dimensionality and size. In many regimes, we find significant improvement over adiabatic time dependent
density functional theory or second Born nonequilibrium GreenÕs function approximations. We briefly
discuss the reasons for the residual discrepancies, and directions for future work.
\end{abstract}

\maketitle
Hybrid methods are a valuable option in physics, to merge 
concepts and perspectives into a more general and effective level of description. 
This work adds an item from condensed matter physics to the list; 
we propose a hybrid method which combines
non-perturbative exchange-correlation (XC) potentials from Time Dependent Density Functional Theory (TDDFT) \cite{GR84,UllrichBook,Botti}  with
many-body perturbative self-energy schemes from Non-Equilibrium Green's Functions (NEGF) \cite{KBE,Keldysh,StefLeeu,BalzBon}, to
deal with systems with strong electronic correlations and out of equilibrium.

An accurate first-principles description of the real-time dynamics of systems with strong electronic correlations
is an important, difficult and basically unsolved problem of condensed matter research. General
frameworks like TDDFT and NEGF do indeed allow for an in-principle-exact treatment
of strong electronic correlations.  However, they both rely on key ingredients [the exchange-correlation (XC)  potential for TDDFT
and the self-energy $\Sigma$ for the NEGF] that in general are only approximately known.
For TDDFT, a systematic and general way to include non-local, non-adiabatic effects 
in the XC potential is lacking, while for NEGF a main
hindrance is that self-energies based on many-body perturbation theory, already computationally 
demanding, are usually inadequate for strong electronic correlations. While
considerable progress has been made for model system far away from equilibrium
(see e.g. \cite{
CapelleUllrich,RubioFuksTokatly,
Maitra15,BalzerEckstein,Werner,Godby,Romaniello}) or
for the {\it ab initio} description of near-equilibrium situations (see e.g. \cite{SharmaGross,Turkovsky}), 
a reliable first-principles treatment of the far-from-equilibrium regime is still lacking.

Here, we suggest a step towards the solution of this problem,  
by a novel combination of TDDFT and NEGF, where 
perturbative (but systematic) memory-effect
corrections augment a non-perturbative local adiabatic treatment of electronic correlations.
The approach is fully conserving in the Kadanoff-Baym sense \cite{conserv}  and, using
the so-called generalized Kadanoff-Baym ansatz \cite{Lipavsky} (see below), 
can be made viable for realistic systems.

Putting in practice our proposal at the {\it ab-initio} level 
requires access to continuum non-perturbative XC potentials, and this point is
addressed at the end of the paper.  However, the scope of our method can already be
illustrated here using simple lattice models. This has the advantage of avoiding complex
implementations and technicalities that, indispensable to deal with real-world systems, are usually
unnecessary (possibly even unwanted) for an explorative assessment of a new
methodology. Our results show that in many situations (see also the supplementary material, SM) the hybrid method
provides significant progress over adiabatic-TDDFT and perturbative schemes for NEGF,
thus holding promise for an improved treatment of the nonequilibrium dynamics of realistic correlated systems.

\indent{\it Models systems.- }
We consider small Hubbard-type 1D and 3D clusters, isolated or coupled to two 1D semi-infinite non-interacting leads. In the latter case, the cluster consists of 1 site (single impurity).
These systems are exposed to time-dependent (TD) local perturbations and/or (where applicable)
to electric biases in the leads. The Hamiltonian for the above setups is
\begin{eqnarray}
\Hh = \Hh _{c} + \Hh _{l} + \Hh _{cl},\label{Ham}
\end{eqnarray} 
which has contributions from the cluster, the leads, and the cluster-leads couplings. In standard notation,
\begin{eqnarray}
 \Hh _{c} = -V'\!\!\!\!\!\sum _{\langle ij \rangle \in C, \sigma} \ch_{i\sigma} ^\dagger \ch_{j\sigma} + 
 \sum _{i \in C} \epsilon _{i} (t) \hat{n}_{i} + 
 \sum _{i \in C} U_i \hat{n}_{i\uparrow} \hat{n}_{i\downarrow},
\label{central}
\end{eqnarray}
where $\langle ij\rangle$ labels nearest-neighbour sites in the cluster $C$, $V'>0$ is the tunneling amplitude,
$\epsilon _i(t)$ are time-dependent on-site energies in the cluster, and $U_i$ are contact-interaction strengths. Further, $\nh_i = \nh_{i\uparrow} + \nh_{i\downarrow}$. For the lead Hamiltonian, $\Hh_{l}=\sum_\alpha \Hh_{\alpha}$, where $\alpha= R (L)$ refers to the right (left) lead, and
\begin{align}
 \Hh_{\alpha } = -V \sum _{\langle ij \rangle \in \alpha,  \sigma} \ch_{i\sigma} ^\dagger \ch_{j\sigma} + 
 \sum _\alpha b_\alpha (t) \hat{N}_\alpha.\label{Hamlead}
\end{align}
Here, $b_\alpha(t)$ is the (site-independent) bias in lead $\alpha$, 
$V > 0$ the tunnelling amplitude and $\hat{N}_\alpha = \sum _{i \in \alpha} \nh _i$.  The coupling between the leads and the cluster (impurity) are given by
\begin{align}
 \Hh_{cl}= -V_{\rm link.} \sum_{\sigma}(\chd _{1_L,\sigma} \ch_{1_C,\sigma} +\chd _{1_R,\sigma} \ch_{1_C,\sigma})+h.c.
\label{leadclus}
\end{align}
All energies units are expressed in terms of the hopping parameter $V'$ (for the 1-site impurity cluster we use $V$ instead),  
and time is measured in the units of the inverse hopping parameter (assuming atomic units). 
We now switch to continuum variables for generality and notational convenience, and provide
some elements of TDDFT and NEGF relevant to our approach.

%
%
%
%
\begin{figure}
\begin{center}
\includegraphics[width=8.7cm]{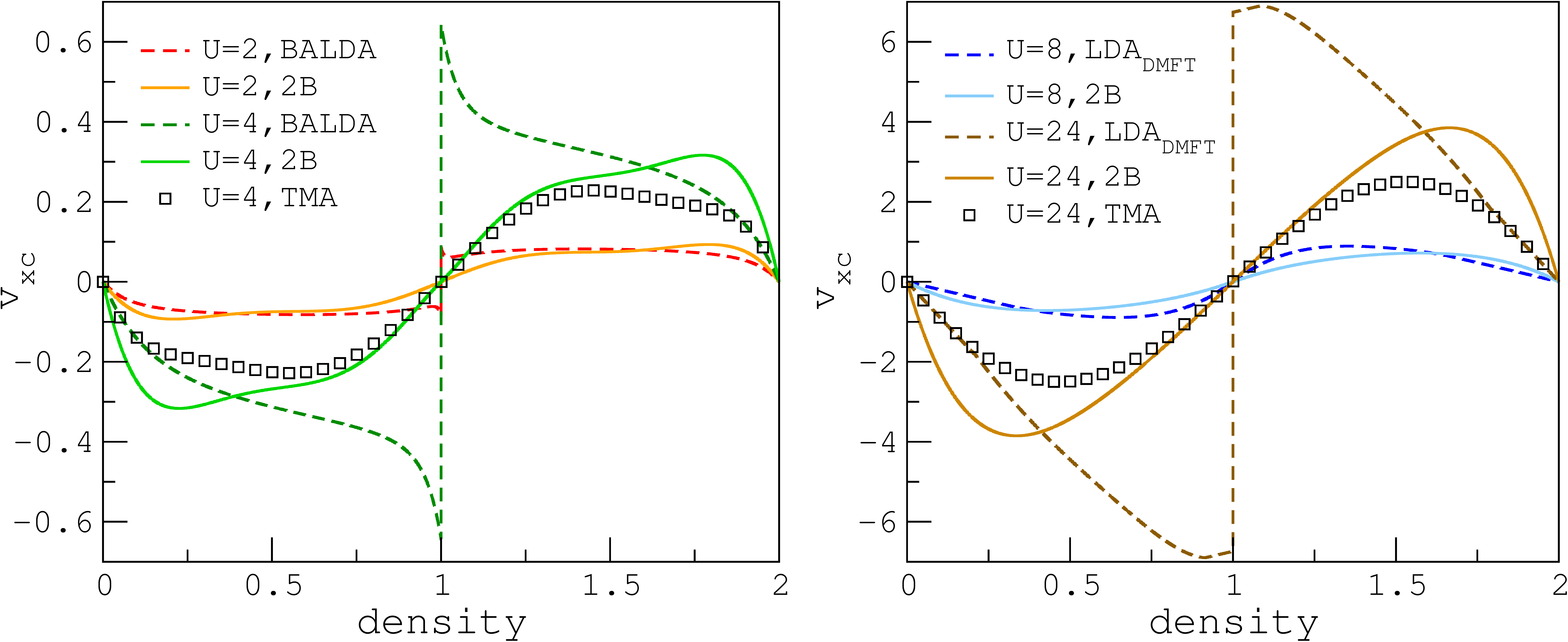}
\caption{XC potentials from the 1D- (left) and 3D- (right) homogeneous Hubbard model.} 
\label{vxc}
\end{center}
\vspace{-0.8cm}
\end{figure}
%
%
%
%
%
{\it NEGF.-} The nonequilibrium propagator $G(1,2)\equiv G(t_{1},\mathbf{r}_{1},t_{2},\mathbf{r}_{2})$ satisfies the equation of motion $[{\rm i}\partial_{t_{1}}-h(1)]G(1,2)=\delta(1,2)+\int_{\gamma}\Sigma (1,3)G(3,2)d3$ (and a similar one for $t_{2}$). Here, $h=T+v_{H}+v_{ext}$ is the single-particle Hamiltonian, with kinetic energy $T$, Hartree potential $v_{H}$,
and external potential $v_{ext}$. $\Sigma = \Sigma_{emb} + \Sigma_{xc} [G]$ is the self-energy, which introduces a memory dependence. We integrate over the Keldysh contour $\gamma$ \cite{KBE,Keldysh}. $\Sigma_{emb}$ is an embedding self-energy which accounts for the leads (if present), 
while $\Sigma_{xc}$ accounts for XC effects \cite{Myohanen2008}.
Standard approximations for $\Sigma_{xc}$ are Second-Born (2BA), T-matrix (TMA) and screened  interaction (GW) \cite{StefLeeu,BalzBon}. For real time, the lesser part of $G$  (denoted $G^{<}$) gives the density $n(t,\mathbf{r})=-{\rm i}G^{<}(t,\mathbf{r},t,\mathbf{r})$ and the current.\\
\indent{\it TDDFT.-} The time-dependent density $n_{KS}$ is obtained in terms of the Kohn-Sham (KS) orbitals $\phi_{\kappa}(t,\mathbf{r})$. These obey the KS equation $[T+v_{KS}(t,\mathbf{r})]\phi_{\kappa}(t,\mathbf{r})={\rm i}\partial_{t}\phi_{\kappa}(t,\mathbf{r})$, where
$v_{KS}=v_{H}+v_{ext}+v_{xc}$, and $v_{xc}$ accounts for XC effects. Then, $n_{KS}(t,\mathbf{r})=\sum_{\kappa}^{occ.}|\phi_{\kappa}(t,\mathbf{r})|^2$. Within a NEGF treatment, the KS density can be obtained from $[{\rm i}\partial_{t_{1}}-h(1)-v_{xc}(1)]G_{KS}(1,2)=\delta(1,2)$, with 
$n_{KS}(t,\mathbf{r})=-{\rm i}G_{KS}^{<}(t,\mathbf{r},t,\mathbf{r})$. In practical implementations, the functional 
dependence of $v_{xc}$ on $n$ is often replaced by an Adiabatic Local Density Approximation (ALDA), i.e. $v_{xc}([n],{\bf r},t) \approx v^{ref}_{xc}(n({\bf r},t))$.\\
\indent{\it A hybrid TDDFT-NEGF approach.-} 
Our proposal is to augment a perturbative self-energy $\Sigma_{xc}^{PT}$ from 
a conserving many-body scheme with a non-perturbative XC potential $v^{np}_{xc}$, local in space/time.
Alternatively, this prescription 
can be be seen as recasting  an ALDA-TDDFT based  on $v^{np}_{xc}$ in a NEGF approach, but augmenting it with a non-local, non-adiabatic
perturbative self-energy $\Sigma^{PT}_{xc}$. 
To avoid double counting we subtract an ALDA potential $v^{PT}_{xc}$
obtained from the same approximation as was used for $\Sigma^{PT}_{xc}$.
The basic equation of our approach is
\begin{align}
\label{kbe}
&[~{\rm i}\partial_{t_{1}}-h(1)-v^{np}_{xc}(n(1))+v^{PT}_{xc}(n(1))\;]~G(1,2) \nonumber
\\&~~~~~~~~~~~~~~~~~~~~=\delta(1,2)+\int_{\gamma}\Sigma^{PT}_{xc}(1,3)G(3,2)d3.
\end{align}

To actually proceed with Eq.(\ref{kbe}), at $t=0$ we solve for $G$, i.e. we find $v^{np}_{xc}[G]$, $v^{PT}_{xc}[G]$ and $\Sigma^{PT}_{xc}[G]$ self-consistently on the imaginary-time track; then we propagate $G$ self-consistently
on the Keldysh contour, thus fulfilling the conservations laws of Kadanoff and Baym.
The hybrid scheme involves no additional computational costs compared to standard NEGF time propagation. 
Since the augmentation $v^{np}_{xc}(t)-v^{PT}_{xc}(t)$ is of the form of a time-local potential, our scheme can similarly be implemented in a density matrix formalism. This means that a Generalized Kadanoff-Baym Ansatz (GKBA) \cite{Lipavsky,Bonitz,Stefanucci} can be employed to reduce computational costs allowing for first-principles calculations of realistic systems. 

\indent{\it The non-perturbative XC potentials-} 
For lattice systems, $v_{xc}^{np}$ depends on the system's dimensionality. 
In 1D, we describe the non-perturbative, adiabatic local correlations in terms of 
$v_{xc}^{np}(t,\mathbf{r})\approx v^{BALDA}_{xc}(n(t,\mathbf{r}))$ \cite{Verdozzi08},
and in 3D in terms of $v_{xc}^{np}(t,\mathbf{r})\approx v^{DMFT}_{xc}(n(t,\mathbf{r}))$ \cite{Karlsson}. 
$v_{xc}^{BALDA}$ is computed with the Bethe-ansatz from the 1D Hubbard model \cite{GunSchonNoack,LimCape}, and $v_{xc}^{DMFT}$ with DMFT \cite{DMFT,DMFT-Hubb.ref} from the 3D homogeneous Hubbard model \cite{Karlsson}.

\indent{\it The $v_{xc}^{PT}$ correction.-} 
For concreteness, in this paper $\Sigma^{PT}_{xc}$ and $v^{PT}_{xc}$ are computed in the
2BA (some results in the TMA are also shown).
The calculation and use of $\Sigma^{2B}_{xc}$ for Hubbard-type
interactions in a NEGF time evolution has been discussed before (see e.g. \cite{Puig}) and is not
repeated here. Rather, we provide additional details of the perturbative correction $v_{xc}^{2B}$.
For the homogeneous (Hubbard) reference system, we use 
$v_{xc}^{2B}(n)=\frac{\partial E_{xc}^{2B}(n)}{\partial n}$, where 
$E^{2B}_{xc}(n)=E^{2B}_{tot.}(n)-T_{0}(n) - E_{H}(n)$, and the 
three terms on the RHS respectively are the total energy in the 
second Born approximation, the non-interacting kinetic energy and the
Hartree energy for the 1D (or 3D) homogeneous Hubbard model. 
We compute $E^{2B}_{tot.}(n)$ in $(\omega,{\bf q})$-space:
\begin{align}
E^{2B}_{tot.}=\frac{-1}{(2\pi)^{D+1}}\int_{-\infty}^{\infty} \! \! \! d\omega \int_{BZ}^{} \! \! d\mathbf{q} 
\ {\rm Im} G^{R}(\omega,\mathbf{q}) f(\omega) (\omega+\epsilon_{\mathbf{q}}),\nonumber
\end{align}
with $G^R$ the retarded propagator, $f$ the statistical Fermi factor (we consider zero temperature), $\epsilon_{\bf{q}}$ the single-particle energies, and 
$ n=\frac{-2}{(2\pi)^{D+1}}\int_{-\infty}^{\infty} \! d\omega \int_{BZ}^{} d\mathbf{q}{\rm Im} G^{R}(\omega,\mathbf{q}) f(\omega)$.
In Fig.~\ref{vxc} we plot $v_{xc}^{2B}$ for the 1D and 3D Hubbard model, for different interaction values.  
We also show the non-perturbative potentials $v_{xc}^{BALDA}, v_{xc}^{DMFT}$ used in Eq. ({\ref{central}).
They exhibit a discontinuity at half-filling, which is always present in 1D but only for large $U$ values in 3D, reflecting  
the Mott-Hubbard metal-insulator transition \cite{Karlsson}. The discontinuity is absent in the 2BA. 
Note that, at exactly half-filling, $v^{np}_{xc}$ and $v_{xc}^{PT}$ are both zero. 
Finally, $v_{xc}^{PT}$ from the TMA is shown. The discontinuity is absent also in this case, and 
 at low/high filling $v_{xc}^{TMA}$ approaches $v_{xc}^{np}$. 

%
%
%
%
\begin{figure}[t]
\includegraphics[width=8.5cm]{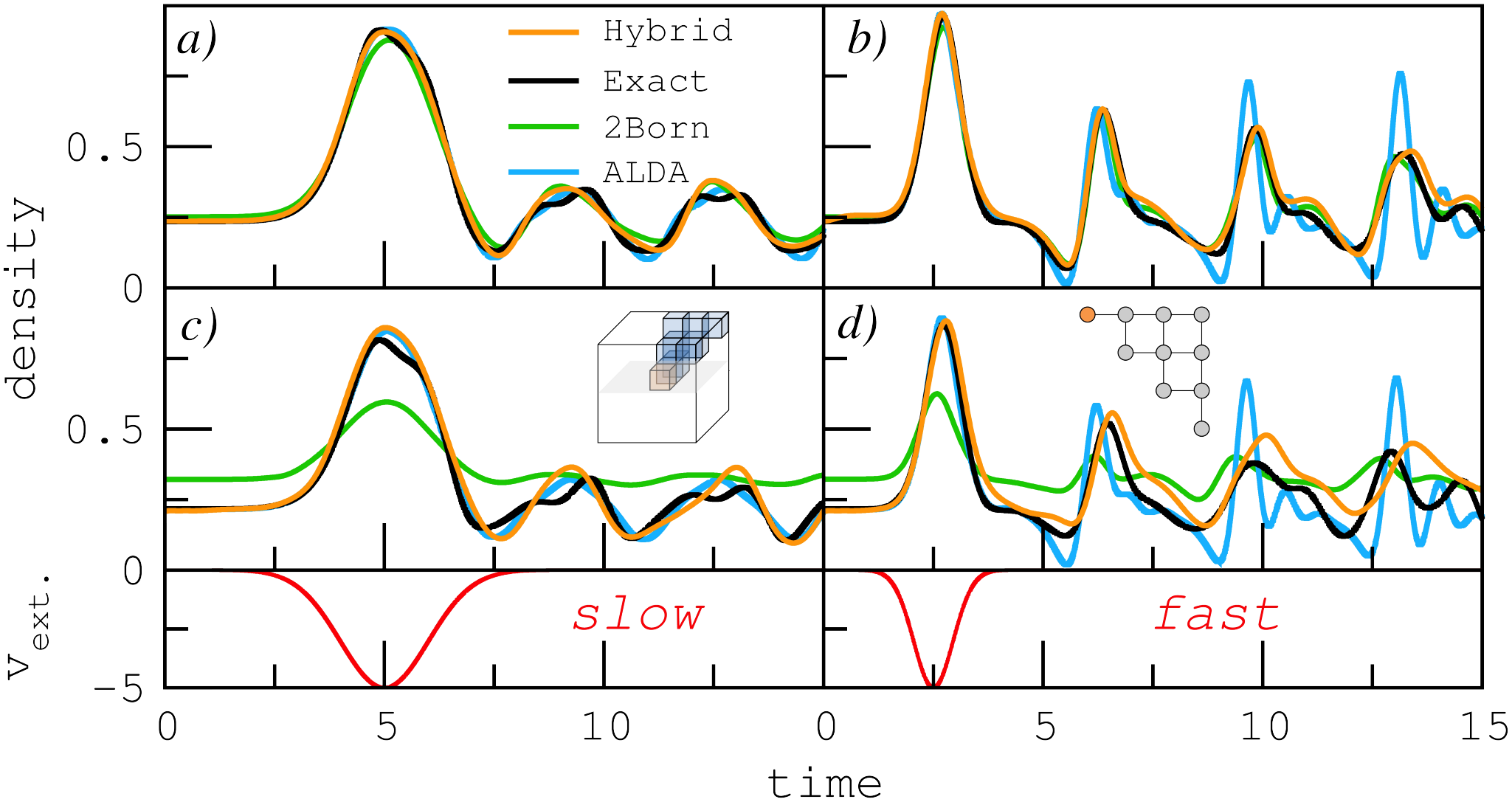}
\caption{Time dependent density at the central site of a $5^3$-site cluster (panel c)
for $U=8$ (top row) and $U=24$ (middle row), for slow (left column) and fast regime (right column). The effective cluster is displayed in panel d. The interacting and perturbed site of the cluster is coloured in orange. The perturbations $V_{ext}$ are shown in the bottom row.}
\label{eqc}
\end{figure}
%
%
%
%

\indent{\it Closed systems: the 3D case. -} 
We start our analysis with a 3D cubic cluster with $5^3$ sites, open boundary conditions, and a single interacting (and perturbed)
site at the cluster centre (Fig.~\ref{eqc}c). We compare time-dependent densities from the hybrid-approach, 2BA and ALDA, against exact results.
The system is highly inhomogeneous, and despite the local character of 
the interaction and external perturbation, non-local effects are important:
the exact $v_{xc}$ (not shown)
can have large nonzero components at all sites \cite{Karlsson}. 
Using symmetry, we map the cluster to
a 10-site one (Fig.~\ref{eqc}d), as in \cite{Karlsson}.
We consider both weak ($U=8$, panels a,b) and strong correlations ($U=24$, panels c,d).
The temporal shape of the external fields we use is Gaussian (bottom-row panels, red curves), 
with a slower or faster onset/offset (in the following, referred to as fast or slow perturbations). For additional time profiles we refer to the SM.

For the weakly correlated, slowly perturbed case (panel a), all approximations follow the exact solution. For the fast perturbation (panel b), non-adiabatic effects emerge, and this leads to the failure of the ALDA; the remaining approximations perform well, with the hybrid method marginally better than 2BA. In contrast, for the slow perturbation and stronger correlations (panel c), the agreement of the 2BA is poor, while the other treatments still follow the exact solution. For the most unfavourable and extreme regime of strong correlations and fast perturbations (panel d), ALDA and 2BA are largely out of  phase, and only the hybrid approximation reproduces the main structures of the exact solution with the correct phase. Overall, the hybrid approximation exhibits a fairly good agreement in all regimes, and is superior to the others in the most extreme regime. 
%
%
%
%
\begin{figure}[t]
\includegraphics[width=8.5cm]{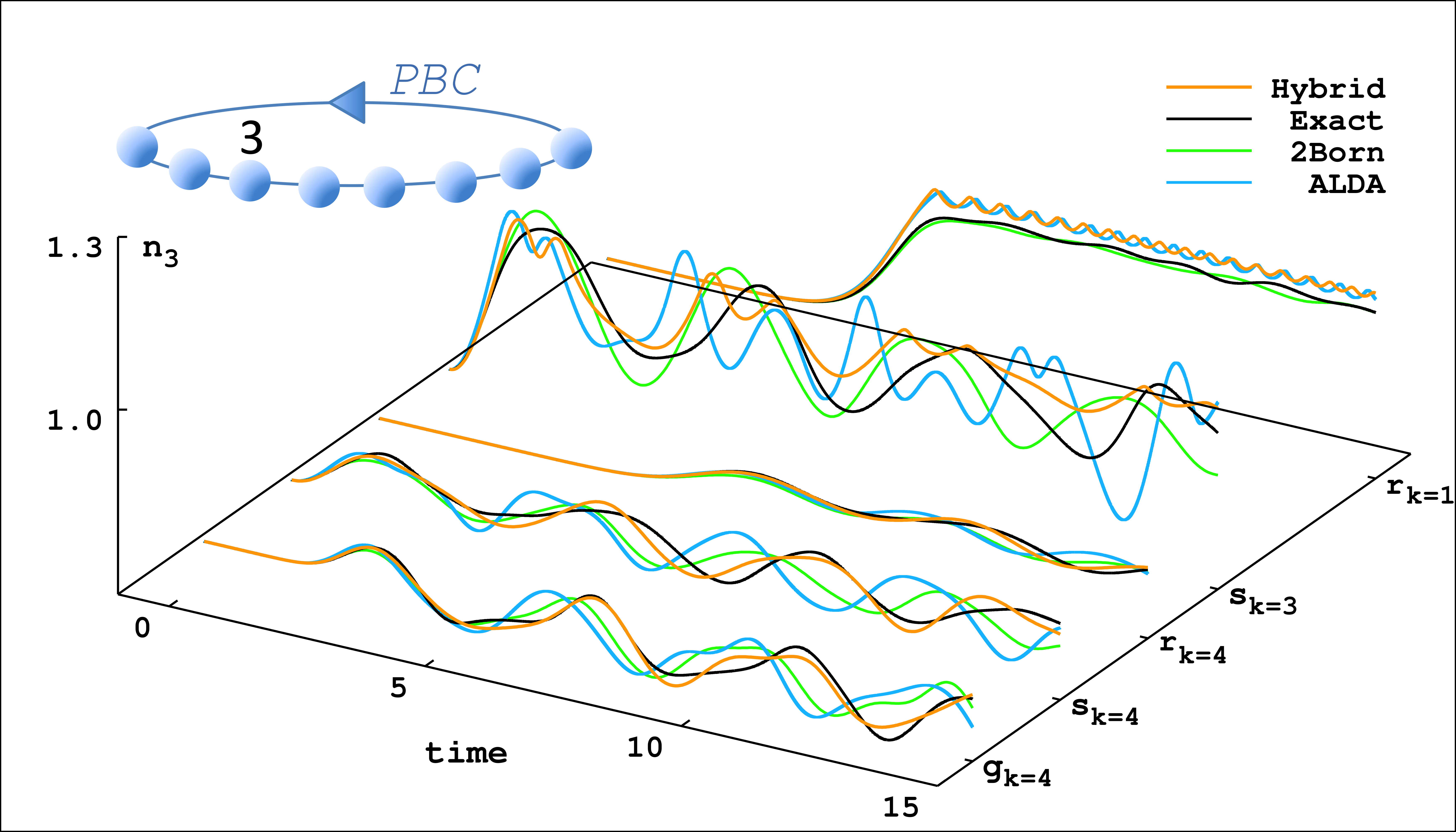}
 \caption{Time evolution of the density $n_3$ at site 3 in a $L=8$  Hubbard ring with $U=4$, under the 
perturbation $V_{ext}(l,t)$. The parameters for the perturbation profile are
$t_0=0, \sigma=1$ for the step (s), $t_0=5.5, \sigma=0.5$ for the ramp (r) and $t_0=2.5, \sigma=\sqrt{0.4}$ for the gaussian (g).
}
\label{cassetto}
\end{figure}
%
%
%
%
%

\indent{\it Closed systems: the 1D case.-} 
We next consider when all sites are interacting and exposed to a space- and time-dependent
perturbation. A 3D system for this situation which is also an exactly solvable benchmark is not easily accessible,
due to the unfavourable scaling of the configuration space. We thus turn to a numerically more convenient 1D test-case 
(this also makes possible to assess the hybrid approach at low-dimensionality), choosing a 1D ring with 8 interacting sites (Fig.~\ref{cassetto}). 
To explore the role of space inhomogeneity, we resort to a (rather artificial) perturbation sinusoidally modulated in space: $V_{ext}(l,t)= \sin(\frac{2\pi}{\lambda_k}l+\phi_k) ~F\bigl(\frac{t-t_0}{\sigma}\bigr)$,
where $\lambda_k=2^k$ ($k=1,2,...,4$) and $F$ is temporal profile.  The phase $\phi_k$ guarantees that the sine nodes are between sites and the amplitude at site $l=3$ has always the same sign. For the time profile, $F(t)\equiv \theta(t)$ (step, s), $F(t)\equiv1/(1+e^{-t})$ (ramp, r) or $F(t)\equiv exp(-t^2)$ (gaussian, g). Results are shown in Fig.~\ref{cassetto}
(for a more systematic study see the SM).

\indent With highly inhomogeneous fields ($\lambda_1,\lambda_2$) no approximation reproduces the exact dynamics. Moreover for $r_{k=1}$ the hybrid method shows artificial density oscillations. The latter, also present in the the TDDFT-ALDA based on $v_{xc}^{BALDA}$, are induced by the sharp discontinuity in $v_{xc}^{BALDA}$ and are not removed by the 2BA self-energy (thus, non-local, non-adiabatic effects beyond the 2BA should be also taken into account). For more homogeneous fields ($\lambda_3$) the different approximations compare more favourably to the exact dynamics with superiority of the hybrid method. Looking at $s_{k=3}$, the hybrid approximation is in phase with the exact curve but, for densities changing across half-filling, it still exhibits the artificial oscillations (see the SM). Further, ALDA does not perform well, and 2BA tends to be out-of-phase with the exact solution. Finally, for a slowly varying-in-space perturbation ($\lambda_4$) the hybrid approach (in contrast to the other approximations) is in excellent agreement with exact results. This applies for all time profiles $g,s,r$.
%
%
%
%
\begin{figure}
\begin{center}
\includegraphics[width=8.5cm]{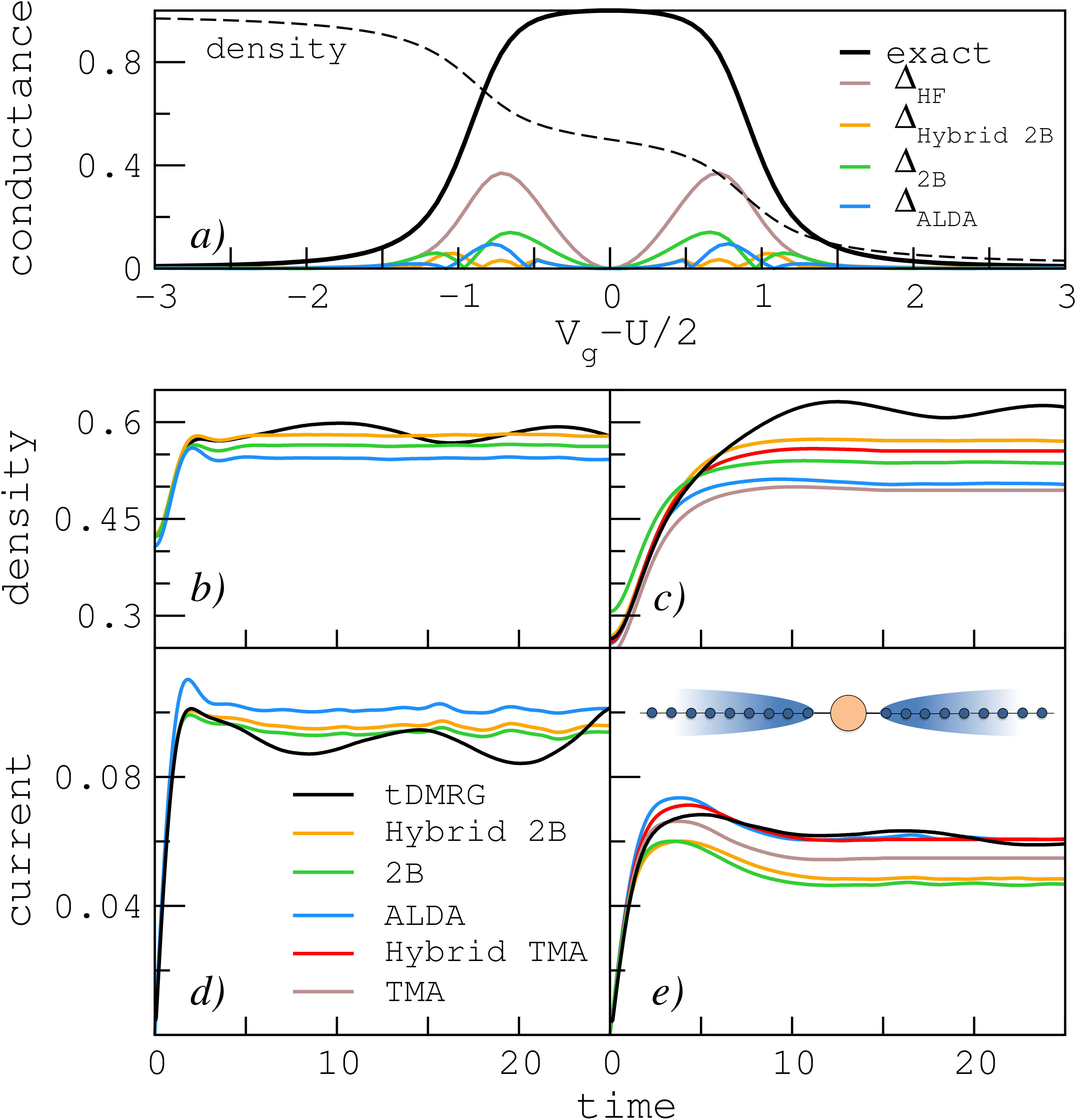}
 \caption{Single-impurity, one-orbital Anderson model with $U=2$ (shown in panel e). a): Linear conductance $G$ in the wide-band-limit for $\Gamma_{WBL}=0.09$ (strong correlations).
The exact $G$ is displayed, together with the Hartree-Fock (HF), 2BA, ALDA and hybrid-method results.
The density/spin-channel $n/2$ at the impurity is also shown (dashed line).  $n/2$ and $G$ share the same vertical scale (in different units). 
 b-e):  Time dependent density $n$ (b, c) and average current $=\frac{j_L+j_R}{2}$ (d, e) for the Anderson impurity with constant impurity gate voltage $V_{gate}=
 \epsilon_0 =0.25$ and bias $b_L(t)=0.5\theta(t)$. The hopping parameter in the leads is $V=1$, the impurity-lead coupling is  $V_{link}=0.5$ (b,d) and $V_{link}=0.3$ (c,e).}  
\label{transport}
\end{center}
\vspace{-0.8cm}
\end{figure}
%
%
%
%

\indent{\it Open systems -} 
Finally, we test the hybrid method in open systems (Fig.~\ref{transport}). Specifically, using
a single-orbital Anderson impurity coupled to two 1D semi-infinite leads \cite{Thygesen} (system shown in Fig.~\ref{transport}e),
we consider
i) the conductance $G$ in the wide-band limit (WBL), Fig.~\ref{transport}a; ii) the finite-bias, finite-lead-width regime,  Fig.~\ref{transport}b-e. 
Starting with i), we find the exact density (and thereby the exact linear conductance via the Friedel sum rule) in the WBL \cite{Wiegmann, Burke,StefanucciKurth,Evers}. Fig.~\ref{transport}a displays for $U=2$ the absolute deviation $\Delta$ from the exact $G$ as function of $V_{gate}$ and for different approximate treatments. We consider stronger correlations ($\Gamma_{WBL}=V^2_{link}/V=0.09$, see the plateau in the conductance); here except for $0.15 < n/2 < 0.28$, the hybrid method performs as the best compared to  2BA or ALDA, and it is significantly better in the range $0.28 < n/2 < 0.42$ (symmetrical considerations apply above half-filling). \\
ii) Next, we consider 1D tight-binding leads (of bandwidth $4V$).  We fix a static $V_{gate}$ to to be away from
the particle-hole symmetric ground state (where $v^{np}_{xc}=v^{PT}_{xc}=0$). As benchmark, we use open-ended, Anderson-impurity finite chains with up to $L=96$ sites treated with tDMRG \cite{tDMRGref,montangero1}. When $V_{link}=0.5$ (panels b,d), the agreement between hybrid and tDMRG densities/currents is fairly good, especially in the transients ($n$ and $j$ from tDMRG never fully reach a steady state within the simulation time, in contrast to hybrid, 2BA, and ALDA. ones However, for stronger correlations and lower transparency $U/V_{link}=2/0.3$ (panels c,e), the impurity density from the hybrid scheme is closest to the tDMRG one than other schemes, whilst for the currents ALDA performs best. 
The unconvincing performance of the hybrid approximation for $U/V_{link}=2/0.3$ comes probably from multiple-scattering processes, neglected by 2BA. 
To corroborate this conjecture we have tested the hybrid method also using the TMA, which includes such processes. In Fig.~\ref{transport}c and e) the TMA hybrid method shows an improvement over the ALDA and the pure TMA calculation and thus supports the conjecture (for an expanded discussion and additional results, see the SM).

{\it Conclusions and outlook.-} By merging elements of TDDFT and NEGF,
we proposed a simple, easy to implement, nonequilibrium scheme aimed
to improve the treatment of local non-perturbative correlation effects and, 
at the same time, to incorporate non-local, non-adiabatic effects. 
Results from Hubbard-type systems are quite encouraging.
Taking a mildly optimistic stand, 
we can argue that our approach extends
the applicability of
ALDA-TDDFT and NEGF based on perturbation theory, thus providing a way forward to
merge (strong) correlations and memory effects in general.
On the other hand, one can certainly envisage situations where non-perturbative {\it and}
non-local correlations are very important, and this is where perhaps
corrections beyond the 2BA (e.g., GWA or TMA or mixed, or other) could be employed. We note
that Hubbard-type systems usually are challenging benchmarks to perturbative
approximations such as 2BA, TMA or GWA. The latter generally perform much
better for continuum systems with long-range interactions.
Thus, we speculatively suggest that 
our  hybrid method could perform even better for
realistic systems. 
This is where the real merits of our proposal 
could possibly be: 
Using continuum XC potentials tailored for strong correlations 
(obtained from e.g. the strictly correlated approach\cite{GoriGiorgi1,GoriGiorgi2,GoriGiorgi3}, where the discontinuities in $v_{xc}$ manifest in a different way) 
and simplifications for perturbative self-energies
(such as the GKBA \cite{Lipavsky,Bonitz,Stefanucci}),
our approach would be a leeway to an improved first principle treatments
of realistic systems in nonequilibrium when strong local electronic correlations and 
memory effects play a role.

\begin{acknowledgments}
We wish to acknowledge M. Puig von Friesen for discussions in the
early stages of this work.
\end{acknowledgments}

\end{document}